\documentclass[11pt, serif]{article}
\pdfoutput=1
\usepackage[margin = 1in]{geometry}
\usepackage{amsmath, amssymb, amsthm, amsfonts}
\usepackage{graphicx}
\usepackage{textcomp}
\usepackage{enumitem}
\usepackage{ragged2e}
\usepackage{subcaption}
\usepackage{pgf,tikz}
\usepackage{libertine}
\usepackage{mathrsfs}
\usepackage{alltt}
\usepackage{float}
\usepackage{wrapfig}
\usepackage[section]{placeins}
\numberwithin{equation}{section}
\numberwithin{figure}{section}
\numberwithin{table}{section}
\usepackage[style=phys, firstinits=true, maxnames=3, backend=bibtex]{biblatex}
\usepackage{hyperref}
\hypersetup{
 colorlinks=true,     %
 urlcolor=blue,       %
 linkcolor=black,     %
 citecolor=red,     %
 plainpages=false,
 breaklinks=true,
 bookmarksnumbered=true,
 bookmarksopen=true,
  pdfpagemode=UseOutlines  %
 }

\def\Title#1{\begin{center} {\LARGE {\textbf{#1}} } \end{center}}
\def\Author#1#2{\begin{center} {\large \textsc{#1}, #2} \end{center}}
\def\Institution#1{\begin{center} {\large {\textit{#1}} } \end{center}}
\def\Abstract#1{\noindent {\normalsize {\bf Abstract:} {\normalfont #1}}}
\def\Conference{\vspace{4mm}\begin{raggedright} {\normalsize {\it Talk presented at the 2019 Meeting of the Division of Particles and Fields of the American Physical Society (DPF2019), July 29--August 2, 2019, Northeastern University, Boston, C1907293.} } \end{raggedright}\vspace{4mm}}

\begin{document}
\Title{COHERENT Plans for D$_2$O at the Spallation Neutron Source}
\Author{Rebecca Rapp}{on behalf of the COHERENT collaboration}
\Institution{Carnegie Mellon University, Pittsburgh, PA 15213, USA}
\Abstract{The Spallation Neutron Source (SNS) is a pulsed source of
  neutrons and, as a byproduct of this operation, an intense source of
  neutrinos via stopped-pion decay.  The COHERENT collaboration uses
  this source to investigate coherent elastic neutrino-nucleus
  scattering (CEvNS) with a suite of detectors.  To enable precise
  cross-section measurements, we must address an estimated 10\%
  uncertainty in our flux calculation associated with the lack of data
  for $\pi^\pm$ production from 1 GeV protons on an Hg target.  We
  present here our Geant4 simulation of neutrino production at the SNS
  and our plans to experimentally normalize this flux with the
  development of a 670 kg D$_2$O detector.  Using the precise cross
  section calculations for neutrino interactions on deuterium, we will
  dramatically reduce our flux uncertainty.}

\Conference
\section{Introduction}
\par Cleanly predicted within the Standard Model (SM) in 1974
\cite{proposition}, coherent elastic neutrino-nucleus scattering
(CEvNS) was first observed in 2017 by the COHERENT collaboration
\cite{sci_mag}.  With the goal of precision CEvNS observations on
multiple targets (CsI, NaI, LAr, Ge), we note that a global and nearly
dominant systematic uncertainty is our understanding of the neutrino
flux from the Spallation Neutron Source (SNS) at Oak Ridge National
Laboratory \cite{arxiv2018}.

\par We use the Geant4 Monte-Carlo framework \cite{geant4} to
calculate the flux of neutrinos in ``Neutrino Alley'', a basement
hallway (8 m.w.e.) where COHERENT deploys our detectors to take
advantage of lower beam-related neutron backgrounds \cite{arxiv2018}.
Section \ref{sec:Sim} describes the simulation, the results we obtain,
and the 10\% uncertainty we place on these results.  Section
\ref{sec:D2O} discusses the proposed development of a D$_2$O detector
that will meet the size constraints of Neutrino Alley, reduce the flux
systematic, and improve COHERENT's physics sensitivity.

\section{Geant4 Simulation of SNS Neutrino Production}
\label{sec:Sim}

\begin{figure}[t]
  \begin{subfigure}[t]{0.565\textwidth}
    \centering
    \includegraphics[width = 0.95\textwidth]{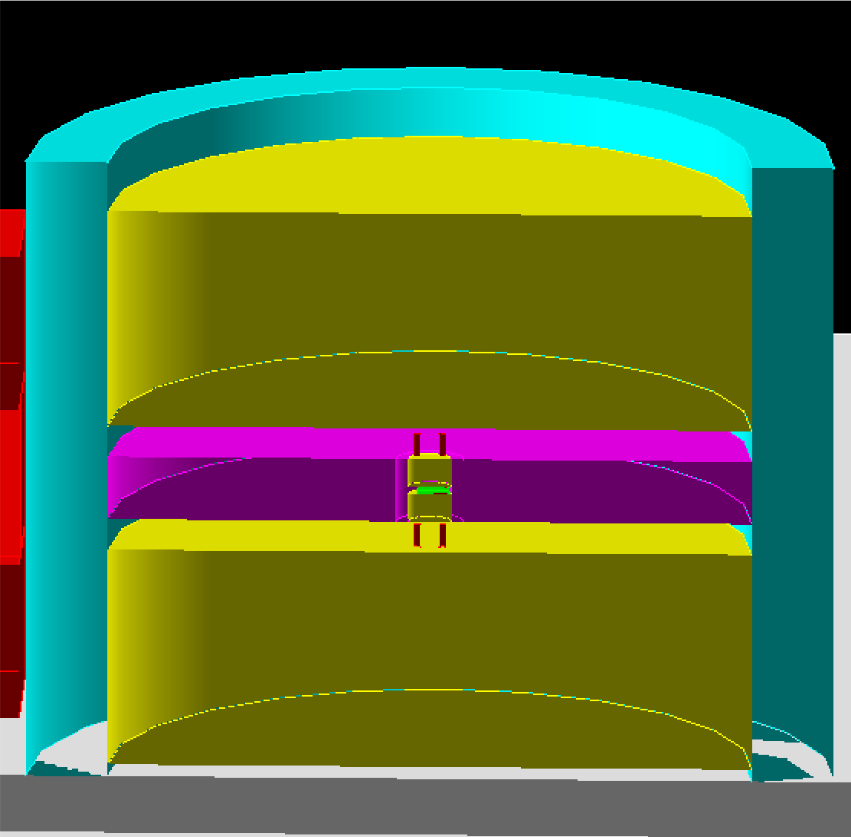}
    \caption{A vertical slice through the layered geometry of the SNS.
      From the outside in: red (mostly off image) is proton beam
      shielding, cyan is outer concrete shielding, large yellow
      cylinders are steel plates, magenta represents the steel
      reflectors, dark red ticks are neutron moderators, small yellow
      cylinders are beryllium plugs, and green is the steel casing of
      the Hg target.}
    \label{fig:geometry}
  \end{subfigure}
  \quad
  \begin{subfigure}[t]{0.425\textwidth}
    \centering
    \includegraphics[width = 0.95\textwidth]{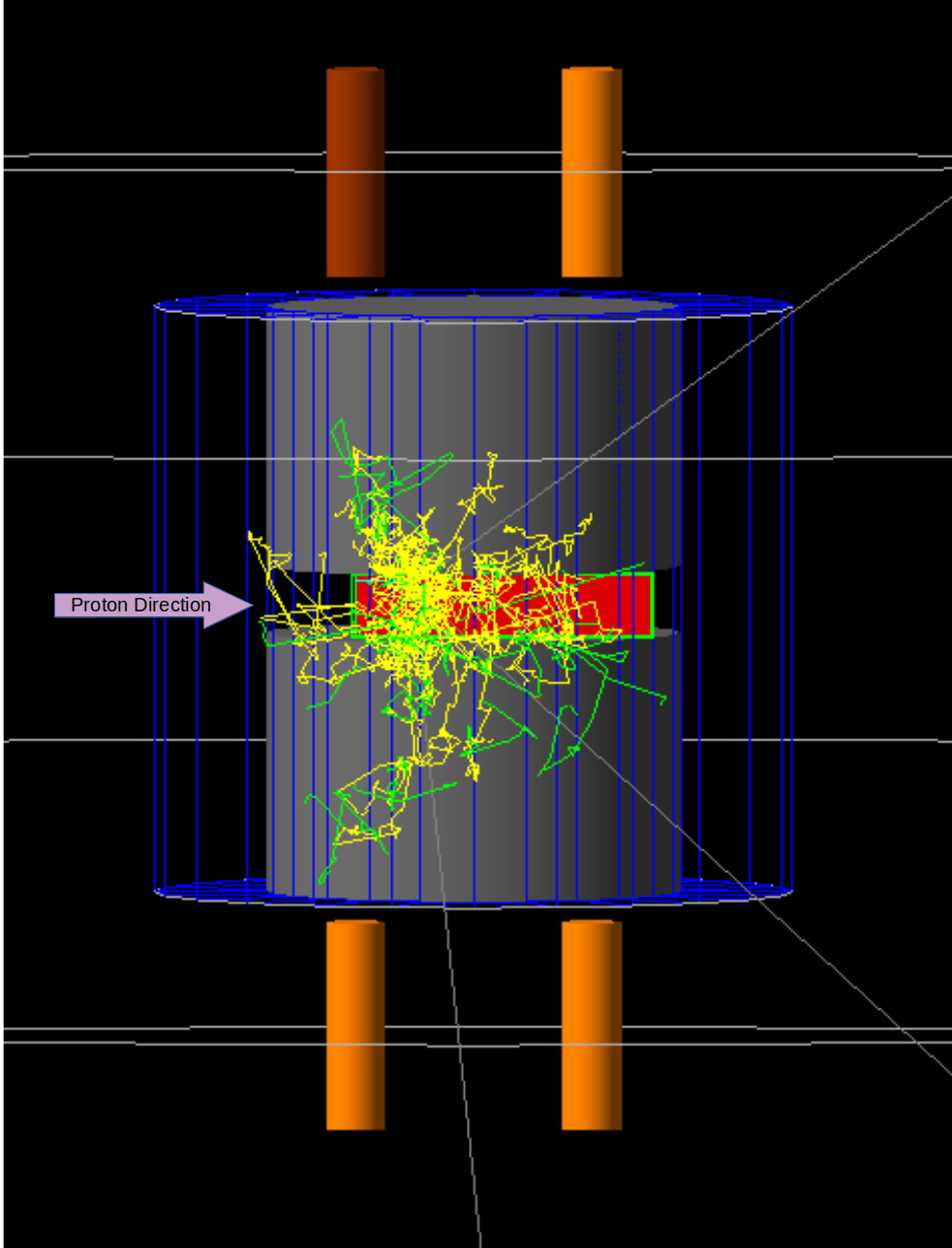}
    \caption{An example Geant4 event, zoomed in on the Hg target
      (red).  In this zoom, we also highlight the (orange) LH$_2$ and
      (brown) H$_2$O moderators, along with the D$_2$O cooling
      (blue).  Protons are uniformly distributed within a centered 70
      cm$^2$ area and generated on the face of the Hg target.}
    \label{fig:event}
  \end{subfigure}
  \caption{Visualizations from the Geant4 simulation of the SNS.}
\end{figure}

\par The SNS accelerates protons to 1 GeV in pulses approximately 700
ns wide, typically 350 ns FWHM.  These pulses strike a liquid mercury
target at a frequency of 60 Hz to create spallation neutrons, but this
process also produces pions which stop in the target and produce
neutrinos from $\pi^+ \rightarrow \mu^+ + \nu_\mu$ and $\mu^+
\rightarrow e^+ + \bar{\nu}_\mu + \nu_e$.  Our simulaton generates
individual, monoenergetic protons at the edge of the target and tracks
the resulting products as they pass through a simplifed Geant4 mockup
of the SNS geometry.  This model was developed in 2015 by the COHERENT
collaboration using technical drawings from ORNL, and Figure
\ref{fig:geometry} illustrates the simplifications.  An example $\nu$
production event is shown in Figure \ref{fig:event}.

\par
\begin{figure}[t]
  \includegraphics[width = 0.495\textwidth]{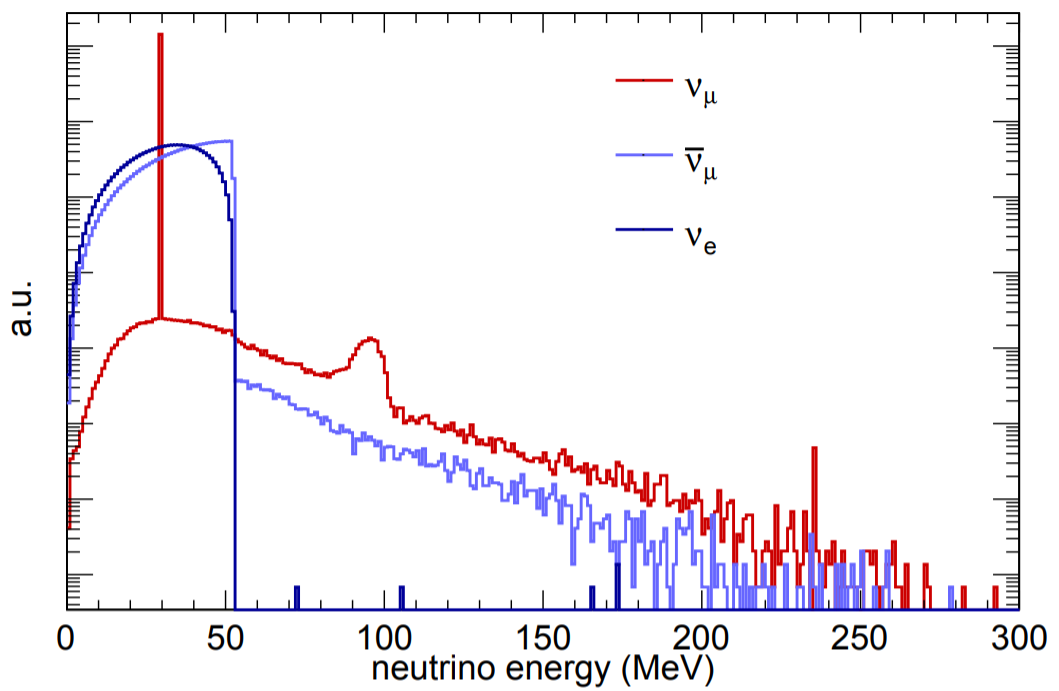}
  \includegraphics[width = 0.495\textwidth]{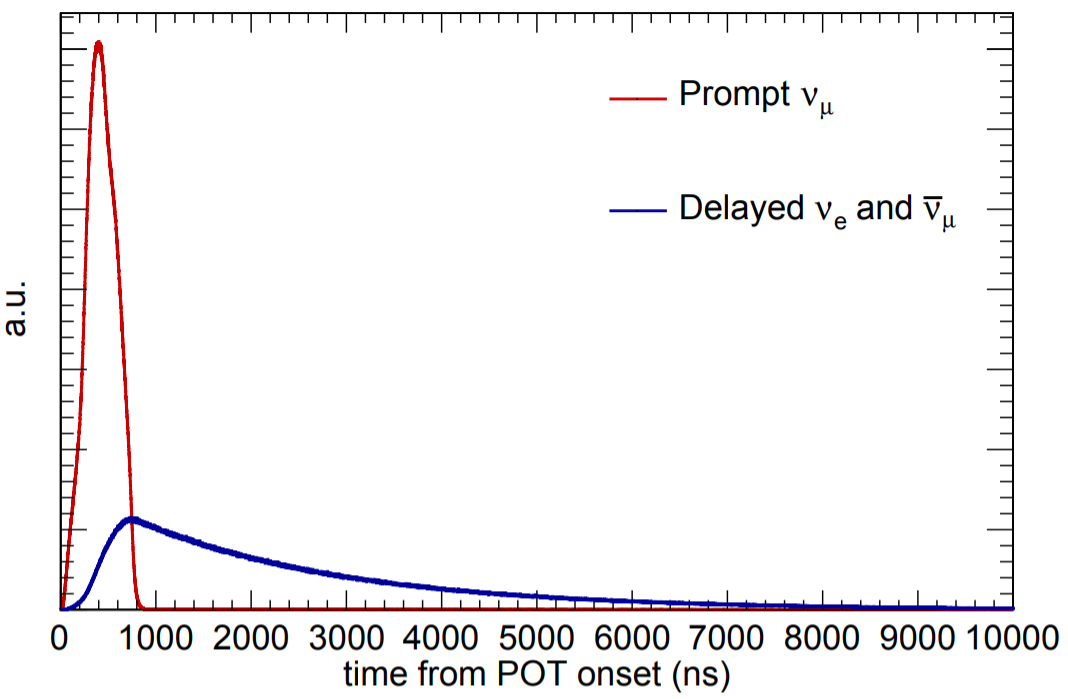}
  \caption{Simulated production energy (left) and timing (right) for
    the dominant $\nu$ flavors at the SNS.}
  \label{fig:spectra}
\end{figure}

 Simulation results for energy and timing information for the
three expected flavors of neutrino are shown in Figure
\ref{fig:spectra}.  In the left hand plot, note the main contributions
come from decay-at-rest processes, resulting in monoenergetic
$\nu_\mu$ from the $\pi^+$ decay and $\nu_e$ and $\bar{\nu}_\mu$
between 0 and 50 MeV from the $\mu^+$ decay.  We do observe
contributions from decay-in-flight $\pi^+$ and $\mu^+$ (the long
tails), the decay of $\pi^-$ leading to $\mu^-$ capture (the peak near
100 MeV), and even some kaon production (monoenergetic peak near 235
MeV), but these contributions are $\sim$1\% or less.  In the right
hand plot, we have convolved the simulation neutrino timing results
(for single protons) with the $\sim$700 ns proton pulse.  We
distinguish between two time regions due to the muon lifetime: the
prompt $\nu_\mu$, and the delayed $\bar{\nu}_\mu$ and $\nu_e$.  Since
the neutrinos are well-timed with respect to the beam, the SNS offers
substantial reduction of steady-state backgrounds.

\par Efforts to validate the simulation indicated that we should use
the QGSP\hspace{-0.05cm}\_\hspace{0.05cm}BERT physics list, which
implements the Bertini intra-nuclear cascade model \cite{bertini}.
Calculations from LAHET \cite{lahet}, a particle transport code that
also implements the Bertini model, are noted to have discrepancies
with available world data, and COHERENT assigns a conservative 10\%
uncertainty to our Geant4 calculations \cite{sci_mag_supplemental}.
With no pion production data from protons at this energy incident on a
mercury target, the best approach for further validation efforts is to
improve the available world data.  We propose to improve the flux
uncertainty using a heavy water detector to normalize the SNS neutrino
flux, as is described in Section \ref{sec:D2O}.

\begin{figure}[b!]
  \centering
  \includegraphics[width = 0.87\textwidth]{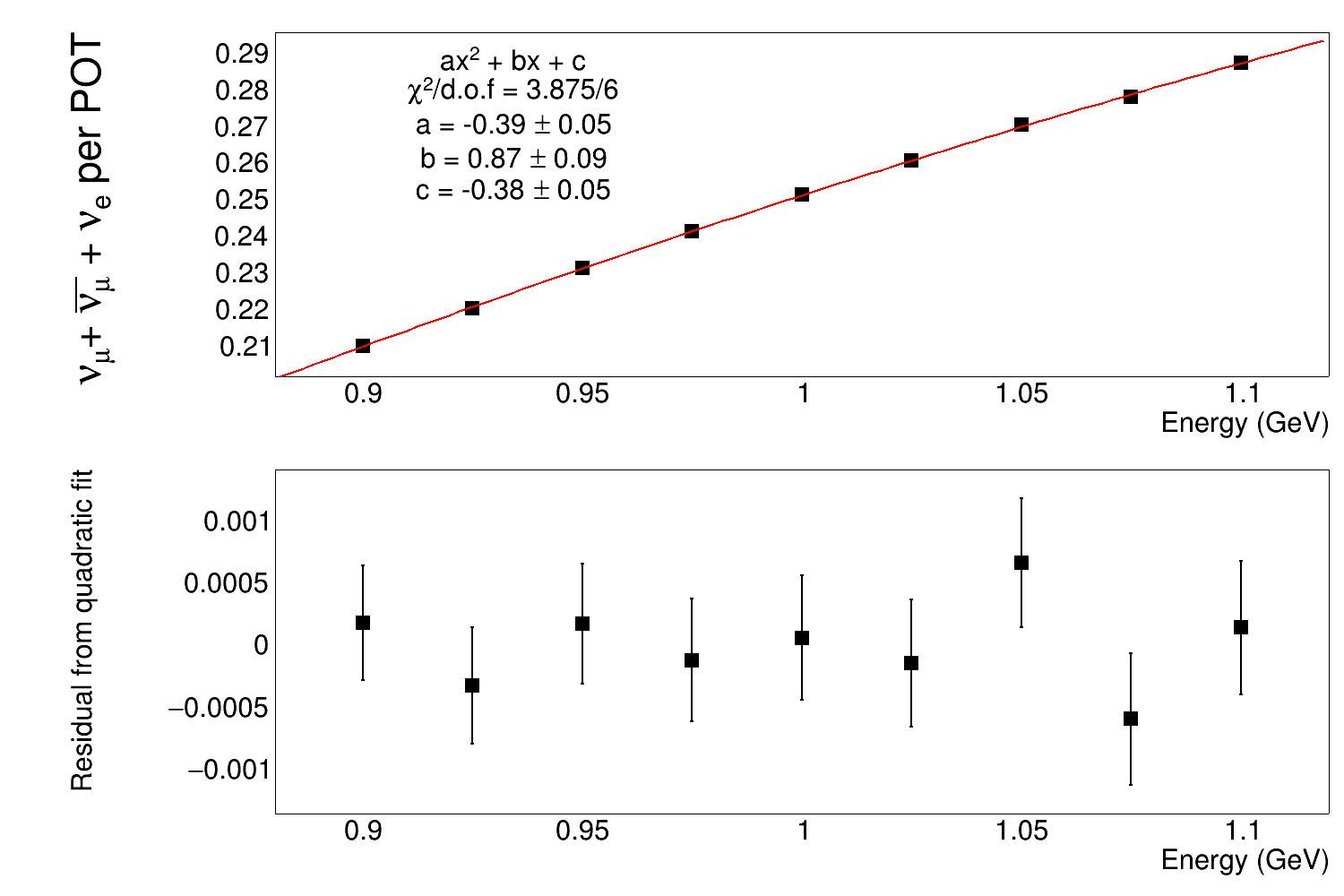}
  \caption{Simulated neutrinos produced per proton on target (POT) as
    a function of proton energy.}
  \label{fig:energyDep}
\end{figure}

\par For the best comparisons to the future results from a D$_2$O
detector, we are also modifying the existing simulation to further our
understanding of the SNS neutrino flux.  We now include information
about a particle's lineage, its production angle, and its creation
position.  These updates inform our efforts as we add features of the
SNS, like the proton beam window, and investigate how decay-in-flight
neutrino production could change the flux at different positions in
Neutrino Alley.

\par Moreover, these simulation updates help to inform the next
generation of experiments as strategies for neutrino physics at the
Second Target Station (STS) start to take shape \cite{sts}.  The only
modification to our simulation involves altering the geometry to the
planned STS target monolith.  The construction of an STS Geant4 model
has begun, and additional features will be added as more details
become available.

\par These updates alone will not drastically reduce our flux
uncertainty.  We do have avenues to compare our simulation results to
existing world data on different targets and proton energies, such as
from HARP \cite{harp}, that could help slightly reduce our
uncertainty, but an experimental normalization of our neutrino flux is
planned to significantly reduce the current 10\% uncertainty (goal of
3\%).

\section{Planned experiment -- D$_2$O detector}
\label{sec:D2O}

\par The SNS produces most of its $\nu_e$ from decay-at-rest $\mu^+$,
meaning that every $\nu_e$ will be accompanied by a $\nu_\mu$ and a
$\bar{\nu}_\mu$.  By measuring the number of $\nu_e$, we can multiply
by a factor of 3 to obtain the total neutrino flux produced by the
SNS.  The deuteron is one of the few nuclei with reasonably calculable
neutrino interactions, and the cross section for $\nu_e + d
\rightarrow p + p + e^-$ has been calculated to 2-3\% uncertainty
\cite{xscnCalc, formaggioZeller, Krasnoyarsk, SNOxscn}.  Thus, the
construction of a heavy water detector will allow us to take advantage
of this well-known cross section and reduce the 10\% systematic on our
neutrino flux.
\begin{figure}[b!]
  \begin{subfigure}[t]{0.56\textwidth}
    \includegraphics[width = \textwidth]{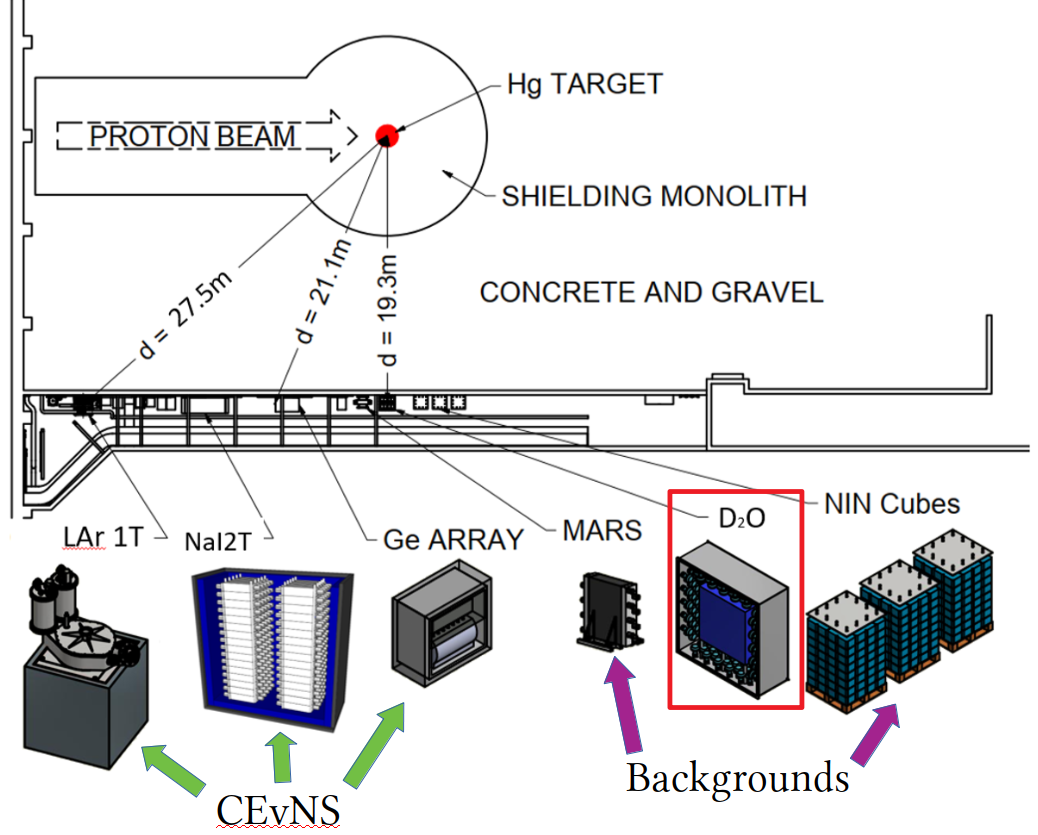}
    \caption{An illustration of the future of Neutrino Alley.  LAr,
      NaI, and Ge will collect CEvNS data, and MARS and the
      Neutrino-Induced-Neutron Cubes will study beam-related neutron
      backgrounds.  The D$_2$O detector will measure the neutrino flux on site.}
    \label{fig:neutAlley}
  \end{subfigure}
  \quad
  \begin{subfigure}[t]{0.43\textwidth}
    \includegraphics[width = \textwidth]{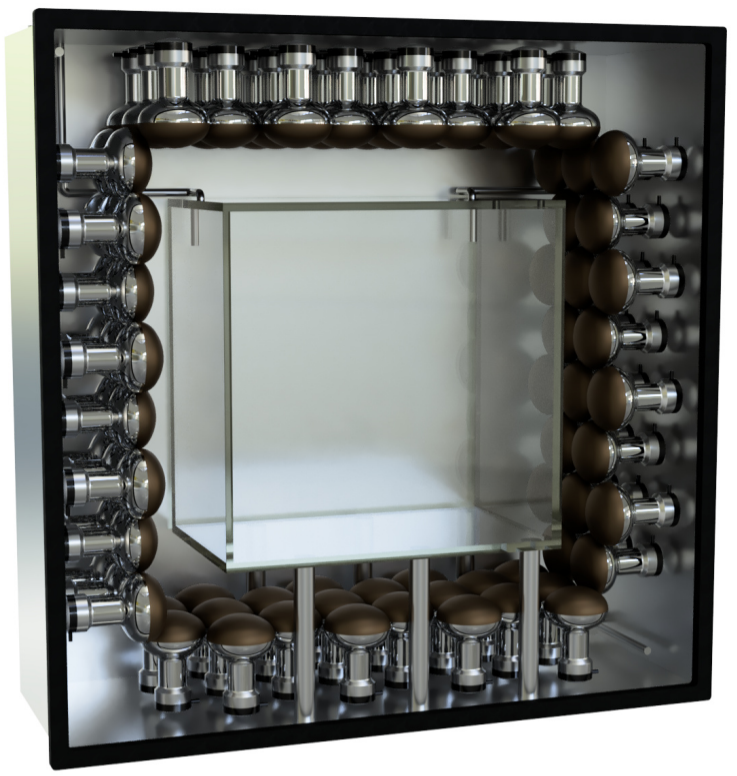}
    \caption{CAD drawing of the planned D$_2$O detector, from Eric Day at CMU.}
    \label{fig:d2oCad}
  \end{subfigure}
  \caption{Illustrations of the planned D$_2$O detector and its placement in Neutrino Alley.}
\end{figure}

\par As we plan out a design for our D$_2$O detector, we are spatially
constrained by our experimental hall, Neutrino Alley (depicted in
Figure \ref{fig:neutAlley}).  Since we occupy an access hallway, we
must always keep 3 ft clear of all instrumentation, meaning any
detector must be smaller than 1 m depth $\times$ 3 m height $\times$ 3
m width.  The largest possible detector to fit within those space
constraints could contain up to 1300 kg of D$_2$O within an inner
acrylic vessel with dimensions 120 cm $\times$ 120 cm $\times$ 60 cm.
We propose a slightly smaller detector, with an inner volume 100 cm
$\times$ 100 cm $\times$ 60 cm, to contain 670 kg of heavy water that
COHERENT has been promised on loan.  Our initial design chooses a
box-like detector for more straightforward construction, and we use
Geant4 to simulate the detector geometry and response.

\par Spatial constraints will prevent photomultiplier tubes (PMTs)
from being placed around the entire detector.  For maximal coverage on
four sides, we would implement 80, 8'' biakali PMTs in staggered rows
around an inner acrylic vessel containing the D$_2$O.  The PMTs are
planned to be submerged in a 10 cm H$_2$O tail-catcher region to aid
in energy reconstruction.  The two non-PMT sides will be fitted with a
Teflon reflector to increase light collection.

\par We expect to have backgrounds from beam-related neutrons, and we
will have shielding around the detector to reduce this.  Cosmic
backgrounds will be monitored with two layers of muon vetos on all
detector sides.  Any signals of neutrino interactions on oxygen are
also background for the $\nu_e + d$ experiment, as the cross-sections
for neutrino interactions on oxygen are not as well known, but we
plan a measurement of the $\nu_e + ^{16}\hspace{-0.08cm}\textrm{O}$
cross-section as a secondary goal of this detector.

\par
\begin{figure}
  \centering
  \includegraphics[width = 0.8\textwidth]{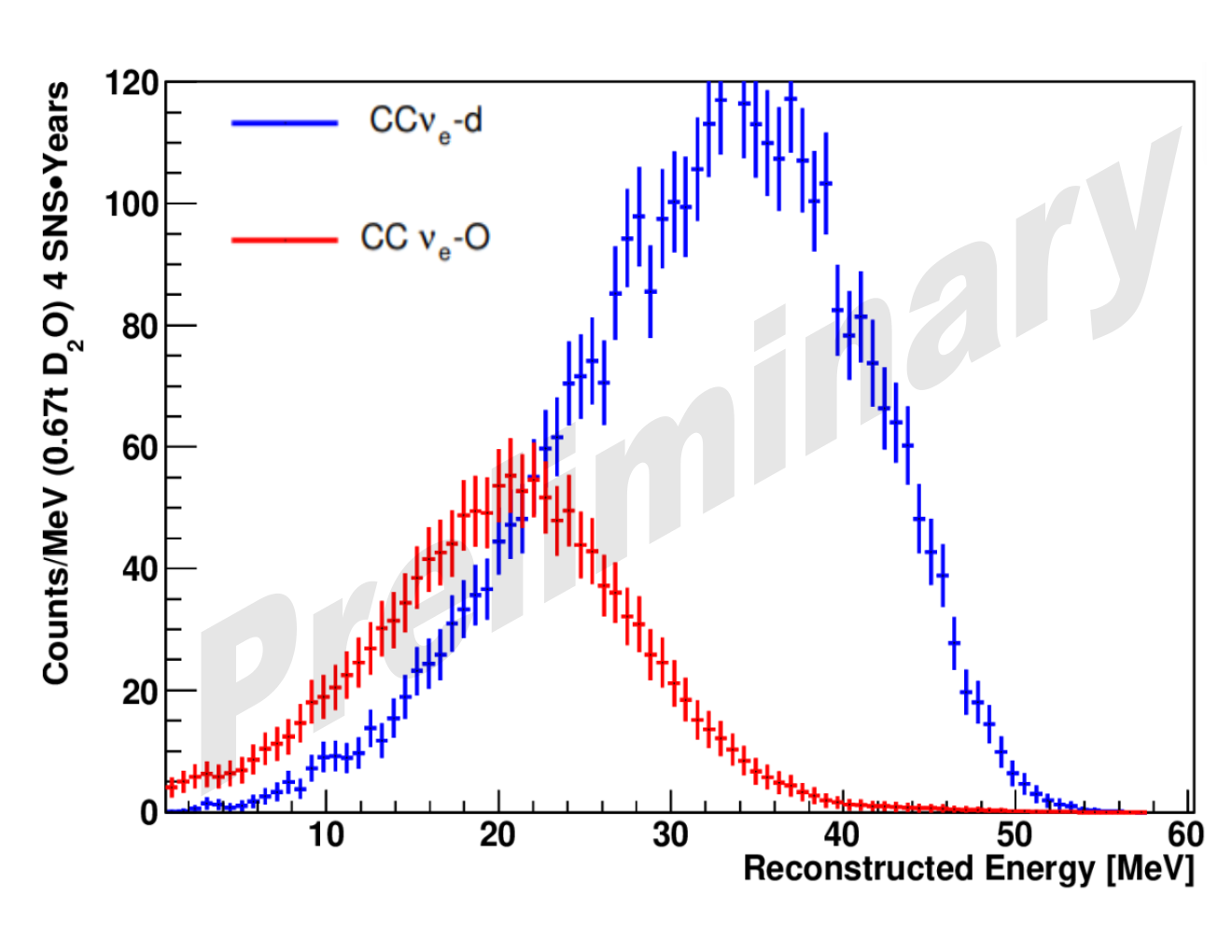}
  \caption{Simulated energy reconstruction with 10 cm H$_2$O after 4
    years collecting data.}
  \label{fig:jason}
\end{figure}

 For the design presented here, we find that four years of
run-time will get us to the percent-level statistical precision to
significantly improve our physics goals.  Our energy reconstruction
comes from the total light collection of the electron created in the
neutrino interaction.  Figure \ref{fig:jason} illustrates the
separation of the deuteron and oxygen components in our predicted
signal after four years of run-time.

\par
\begin{figure}
  \begin{subfigure}[t]{0.495\textwidth}
    \includegraphics[width = \textwidth]{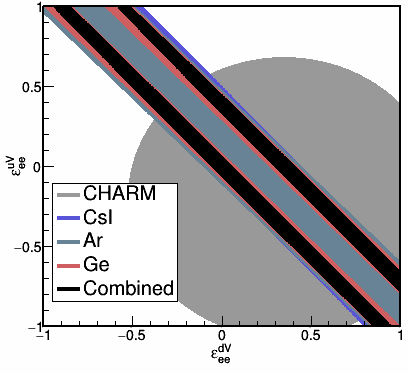}
    \caption{10\% flux uncertainty.}
    \label{fig:nsi_nod2o}
  \end{subfigure}
  \quad
  \begin{subfigure}[t]{0.495\textwidth}
    \includegraphics[width = \textwidth]{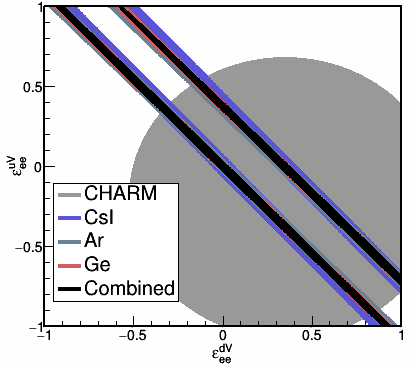}
    \caption{3\% flux uncertainty.}
    \label{fig:nsi_d2o}
  \end{subfigure}
  \caption{Predicted constraints on neutrino-quark interactions beyond
    the Standard Model with planned COHERENT detectors.}
  \label{fig:nsi}
\end{figure}

 As soon as D$_2$O starts taking data, we will begin to reduce our
flux uncertainty and can adjust the Geant4 models.  We aim to reduce
our flux systematic to 3\% uncertainty with this experimental
normalization, and Figure \ref{fig:nsi} is an example of how this will
affect COHERENT's physics sensitivity.  Figure \ref{fig:nsi_nod2o}
plots our predicted constraints on neutrino-quark interactions beyond
the Standard Model (BSM) with our current flux systematic of 10\%.
Using the goal of a 3\% flux systematic, Figure \ref{fig:nsi_d2o}
shows the improved constraints on this parameter space.

\par As we explore funding avenues for the development of this
detector, we are working to optimize the detector design through
simulation.  Design decisions under investigation include the overall
geometry (box-like vs. cylindrical), PMT placement (what is the fewest
number of PMTs needed for total light coverage), the use of Teflon
reflectors (can we reconstruct Cherenkov rings), and the material in
our tail-catcher region (H$_2$O or mineral oil).  In order to validate
these simulations, a small prototype is under development and will use
up to four PMTs.  This prototype will also be used to study backgrounds
in potential locations for the full-scale detector.

\section{Summary}
\par The COHERENT collaboration is moving towards precision
measurements of CEvNS, and must reduce a global systematic on our
understanding of the neutrino flux.  Due to the lack of pion
production data for 1 GeV protons on a liquid mercury target, we
propose the development of a dedicated detector to reduce our
uncertainty to the percent level.  Using a loan of 670 kg of D$_2$O,
we would reduce our current 10\% global flux systematic to the percent
level after four years of run time.  With funding for the largest
possible detector, we would use 1300 kg of D$_2$O and be able to
reduce our flux uncertainty with only two years of run time. This
detector could then also be used to normalize the neutrino flux from
pion production on a tungsten target at the Second Target Station.

\section*{Acknowledgements}
\par The author gratefully acknowledges her support from the
J. Michael McQuade Graduate Fellowship in Physics.  This material is
based upon work supported by the U.S. Department of Energy, Office of
Science, Office of High Energy Physics.  This research used resources
of the Spallation Neutron Source, which is a DOE Office of Science
User Facility.

\clearpage
\printbibliography

\end{document}